\documentclass[sigconf,screen]{acmart}

\def\BibTeX{{\rm B\kern-.05em{\sc i\kern-.025em b}\kern-.08emT\kern-.1667em\lower.7ex\hbox{E}\kern-.125emX}}

\usepackage{amsfonts}
\usepackage{graphicx}
\usepackage{array}
\usepackage{url}
\usepackage{fancybox}
\usepackage{multirow}
\usepackage{color}
\usepackage{colortbl}
\usepackage{balance}
\usepackage{fancyhdr}
\usepackage{subfig}
\usepackage{diagbox}
\usepackage{bbding}
\usepackage{placeins}
\usepackage{booktabs}
\usepackage{enumitem}
\usepackage{xcolor,lipsum}
\usepackage{textcomp}
\usepackage{algorithmic}
\usepackage{listings}
\usepackage{color}
\usepackage{commath}
\usepackage{comment}
\usepackage{bm}
\usepackage{microtype}

\definecolor{light-gray}{rgb}{.906,  .902,  .902}

\setcopyright{acmlicensed}
\acmDOI{10.1145/3663529.3663850}
\acmYear{2024}
\copyrightyear{2024}
\acmSubmissionID{fsecomp24industry-p68-p}
\acmISBN{979-8-4007-0658-5/24/07}
\acmConference[FSE Companion '24]{Companion Proceedings of the 32nd ACM International Conference on the Foundations of Software Engineering}{July 15--19, 2024}{Porto de Galinhas, Brazil}
\acmBooktitle{Companion Proceedings of the 32nd ACM International Conference on the Foundations of Software Engineering (FSE Companion '24), July 15--19, 2024, Porto de Galinhas, Brazil}
\received{2024-02-08}
\received[accepted]{2024-04-18}


\begin{document}
\title[Easy over Hard: A Simple Baseline for Test Failures Causes Prediction]{Easy over Hard: A Simple Baseline for Test \\ Failures Causes Prediction}

% \keywords{TODO comment, Obsolete comment, Code-Comment Inconsistency, Code-comment co-evolution, BERT model}
% \renewcommand{\shortauthors}{Gao et al.}
%\include{} will open a new paper to display the words, while input does not.

\author{Zhipeng Gao}
\affiliation{%
\institution{Zhejiang University}
\country{China}
}
\email{zhipeng.gao@zju.edu.cn}

\author{Zhipeng Xue}
% \authornote{Zhejiang University}
\affiliation{%
\institution{Zhejiang University}
\country{China}
}  
\email{zhipengxue@zju.edu.cn}

\author{Xing Hu}
\authornote{This is the corresponding author}
\affiliation{%
\institution{Zhejiang University}
\country{China}
}
\email{xinghu@zju.edu.cn}

\author{Weiyi Shang}
\affiliation{%
\institution{University of Waterloo}
\country{Canada}
}
\email{wshang@uwaterloo.ca}

\author{Xin Xia}
\affiliation{
\institution{Huawei}
\country{China}
}
\email{xin.xia@acm.org}


\begin{abstract}
The test failure causes analysis is critical since it determines the subsequent way of handling different types of bugs, which is the prerequisite to get the bugs properly analyzed and fixed. 
After a test case fails, software testers have to inspect the test execution logs line by line to identify its root cause. 
However, manual root cause determination is often tedious and time-consuming, which can cost 30-40\% of the time needed to fix a problem.  
Therefore, there is a need for automatically predicting the test failure causes to lighten the burden of software testers. 
In this paper, we present a simple but hard-to-beat approach, named {\sc NCChecker} (\underline{N}aive Failure \underline{C}ause \underline{C}hecker), to automatically identify the failure causes for failed test logs.
Our approach can help developers efficiently identify the test failure causes, and flag the most probable log lines of indicating the root causes for investigation. 
Our approach has three main stages: log abstraction, lookup table construction, and failure causes prediction. 
We first perform log abstraction to parse the unstructured log messages into structured log events. 
{\sc NCChecker} then automatically maintains and updates a lookup table via employing our heuristic rules, which record the matching score between different log events and test failure causes. 
When it comes to the failure cause prediction stage, for a newly generated failed test log, {\sc NCChecker} can easily infer its failed reason by checking out the associated log events' scores from the lookup table. 
We have developed a prototype and evaluated our tool on a real-world industrial dataset with more than 10K test logs.
The extensive experiments show the promising performance of our model over a set of benchmarks. 
Moreover, our approach is highly efficient and memory-saving, and can successfully handle the data imbalance problem. 
Considering the effectiveness and simplicity of our approach, we recommend relevant practitioners to adopt our approach as a baseline to beat in the future. 

\end{abstract}

\maketitle

\section{INTRODUCTION}
\label{sec:intro}
% Background of Testing and  Logs
As modern software has become much larger and more complex, software defects and bugs are unavoidable in such systems. 
These software defects can lead to the system failures and its degraded quality (e.g., performance, reliability and/or security)~\cite{he2021survey, oliner2012advances, zhang2019inflection, mi2013toward, qiu2021deep}. 
% why do we need testing 
To minimize the number of delivered errors and mitigate the risk of system failures, developers, and/or testers usually resort to software testing by running test cases.
% the relationship between testing and test logs 
The risk of software failures may be considered low with passing all test cases. 
By contrast, the risk of software failures is significantly high if the test case fails. 
Once a test case fails, it is necessary for developers and/or testers to further investigate the reasons for its failure by analyzing the test execution results, which are typically stored in log files. 

% the importance of logs 
Developers use logs to record valuable runtime information (e.g., important events, program variables values, trace execution, runtime statistics, and even human-readable messages).
The rich and detailed information recorded in log files are considered as the most important and useful resources to help developers and/or testers understand system failures and identify potential failure causes~\cite{miranskyy2016operational, pecchia2015industry, fu2014developers, yuan2012improving, amar2019mining, li2023did}.  
Moreover, because logs are often the only available data that reports the software runtime information, logs are referred to as the most common accessible resources for diagnosing system failures. 

% the importance of failure causes predictions 
There are many causes that can lead to test failures (e.g., environmental condition problem, source code problem, and software version problem).
Different types of failures have their own corresponding action to perform (e.g., submitting bug reports to developers, rerunning test scripts, submitting exception messages to software maintainers). 
Therefore, it is essential to identify the failure causes in a timely manner such that the corresponding experts can be assigned shortly in order to  review, analyze and fix the bugs without further delay. 
On the other hand, the modern software systems grow rapidly and become more mature. Different software, hardware and services are tightly integrated, leading to ever higher difficulty in diagnosing test failures.
Prior work reports that more than 100 billion US dollars has been spent on failure diagnosis process and manual determination of a test failure root causes can consume half of the total time for fixing a software issue~\cite{zhang2019inflection, zawawy2010log}. 
In practice, it is thus preferable to have toolkits that can automatically diagnose the cause of test failures. 
However, making such a tool is difficult due to the following challenges: 
\begin{enumerate}
    \item \textit{\textbf{Dealing with information overload.}} 
    Information overload is a common challenge for different software engineering tasks~\cite{gao2020technical, gao2023know, xu2017answerbot, xue2023acwrecommender, yang2024federated, li2023they, nie2017data}. 
    To find out the causes of the test failures, software testers have to read and digest the test logs carefully. 
    However, these test logs are often too large to examine manually. 
    In a large software company, thousands of test failures are reported daily resulting into a huge amount of logs, with each log containing hundreds of test steps and thousands of log lines~\cite{herzig2015empirically, amar2019mining, xu2009detecting}. 
    For example, in our study, we collected more than 10K logs from our industrial partner. A test log file consists of 3-4K log lines on average and the largest log file contains more than 550K log lines.
    This huge amount of information goes far beyond the level that testers can handle, and it is extremely difficult and inefficient for testers to manually figure out the failure causes~\cite{mariani2008automated, el2020systematic}. 
    To alleviate such problems, practitioners have crafted extensive regular expressions to analyze test failure causes.
    However, the complexity of the runtime behaviors during testing makes the definition and maintenance of such regular expressions a time-consuming and error-prone task; while the performance is still far from satisfactory~\cite{he2016experience, zhu2019tools}.
    % It is necessary to provide a tool for testers to 
    
    \item \textit{\textbf{Dealing with imbalanced datasets.}}
    As mentioned above, the test failures are caused by various number of reasons. 
    However, not all failure causes happen equally. 
    For example, in our study, the vast majority of failure causes are bug-related issues (i.e., 63\%) and environmental problems (i.e., 23\%), the third-party library issue only accounts for a very small proportion of all failure cases (i.e., 1.4\%).
    Despite previous study~\cite{jiang2017causes} has proposed a similarity-based approach to predict the multiple failure causes, the performance of the similarity-based approaches will decrease dramatically on such a highly imbalanced dataset, especially for these minority failure classes. 
    It is thus valuable to have an approach that focuses on few-shot samples and prevents the majority of samples from excessively affecting the learning process.

    \item \textit{\textbf{Dealing with the rapidly increased latency and memory.}} 
   % Another issue concerns with rapidly growth of the number of logs over time. 
    Considering the large-scale software systems run on 24x7 basis, the generated logs are typically huge~\cite{mi2013toward, he2018identifying}. 
    Analyzing the archived logs in such a huge volume brings challenge to the latency and memory usages. 
    For example, the approaches proposed by previous studies~\cite{amar2019mining} need to compare logs against a library of historical referenced logs to identify the test failure causes.
    If the library increases with the velocity of the rapidly accumulated logs (e.g., 50 gigabytes per hour in Google system), simply reading these logs into memory for comparison purposes and retrieving relevant logs can cost significant time.  
    Therefore, dealing with the rapidly increased logs and ever-increasing computation resources is a major challenge. 

\end{enumerate}

To overcome the aforementioned challenges from practice, in this work, we propose {\sc NCChecker} (\textbf{\underline{N}}aive Failure \textbf{\underline{C}}ause \textbf{\underline{Checker}}), a heuristic rule-based approach for failure causes analysis that learns from the large volumes of test logs. 
{\sc NCChecker} is simple and contains three stages: log abstraction, lookup table construction and failure causes prediction. 
In the first stage, we perform log abstraction to convert each unstructured log into structured log events. 
The structured information contains essential of log lines without any noisy details and can then be used as input for the downstream tasks. 
In the second stage, we create the failure reason lookup table by using four simple but effective heuristic rules. 
The rows of the lookup table are different log events abstracted from the first stage, while the columns of the lookup table are different failure causes (i.e., environmental issues, bug related issues, test script issues, and third-party library issues). 
The cell of the lookup table contains the relevant scores estimated by {\sc NCChecker} by using our heuristic rules. 
For a given log event and a failure reason, the higher the score, the more relevant the log event is associated with the particular failure reason. 
When it comes to the last stage of failure causes prediction, for a newly reported failing test case, we parse the test log into a sequence of existing log events. 
Afterwards, for each log event, we check out the scores from the above lookup table for different failure reasons. 
For a given failure reason, we sum up the checked out scores from all the possible log events for this specific failure reason. 
A failure reason with the maximum value will be selected as the final prediction failure cause.

To evaluate {\sc NCChecker}, we collect more than 10K test logs from our industrial partner, which is a leading information and communication technology company. 
The experimental results show that {\sc NCChecker} outperforms several state-of-the-art approaches by a large margin. 
Moreover, {\sc NCChecker} performs well with respect to the imbalanced dataset and can successfully identify the failure causes in minority. 
In addition, {\sc NCChecker} is efficient and consumes low memory for log analysis. 
In summary, this study makes the following contributions:
\begin{enumerate}
    \item We propose a simple but hard-to-beat approach to address the challenges of test failure analysis. 
    {\sc NCChecker} can assist developers and/or testers to correctly and efficiently diagnose test failures. 
    \item We construct a dataset with more than 10K test logs to evaluate and verify the effectiveness of our model. The dateset involves more than 7K failed logs and 3K passed logs in total. 
    The failure causes of these test logs are manually verified by testers. 
    \item We conduct comprehensive experiments to investigate the effectiveness of our approach. 
    The experimental results show that our approach is effective and efficient. 
    \item Considering the effectiveness and simplicity of our approach, we recommend developers to apply our approach in practice and researchers to adopt our approach as a baseline to beat in the future. 
\end{enumerate}

The rest of the paper is organized as follows.
Section~\ref{sec:pre} presents the dataset overview and key insights. 
Section~\ref{sec:approach} presents the details of our approach.
Section~\ref{sec:eval} presents the baseline methods, the evaluation metrics, and the evaluation results.
Section~\ref{sec:related} presents the related work.
Section~\ref{sec:threats} presents the threats to validity. 
Section~\ref{sec:con} concludes the paper with possible future work.

\section{Preliminary}
\label{sec:pre}
\begin{table}%[t]
\caption{An Overview of the Collected Log Datasets }
% \vspace{-5pt}
\label{tab:data_overview}
\begin{center}
\begin{tabular}{llr}
    \toprule
    {\bf Log Type} & {\bf Measurement} & {\bf Value}  \\
\midrule
\multirow{4}{*}{\bf Failed Logs}
    & \# Logs            & 7,159   \\ 
    & Avg. Log Lines     & 3,905   \\ 
    & Max. Log Lines     & 550,732   \\ 
    & Total File size    & 3.2G   \\ \midrule
    \multirow{4}{*}{\bf Passed Logs}
    & \# Logs            & 3,286   \\ 
    & Avg. Log Lines     & 4,564   \\ 
    & Max. Log Lines     & 270,108   \\ 
    & Total File size    & 1.7G   \\ \bottomrule
\end{tabular}
\end{center}
% \vspace{-10pt}
\end{table}

In this section, we first present an overview of our log datasets.
Afterward, we introduce four key insights that guide the design of the heuristic rules of our approach. 

\subsection{Data Overview}
% \subsubsection{Log Data Statistics.}
In this study, we collected 10,445 test logs (including 7,159 failed test logs and 3,286 passed test logs) from our industry partner. 
We counted the log lines and file size of each test log, and the overall data statistics of the dataset are summarized in Table~\ref{tab:data_overview}. 
In our work, each failed test log is manually labeled with a specific test failure cause. 
There are four types of failure causes with respect to our collected test logs. Different types of failure causes are expected to be handled by different kinds of solutions. 
We now describe the details of different failure causes as follows: 

\begin{itemize}
    \item \textbf{Bug related issues (C1):} 
    The bug related issues are concerned with general software system bugs due to coding mistakes, compatibility problems, and security vulnerabilities etc,. % outdated and/or the updated software releases problems.
    When the bug related issues occur, it is necessary to notify the software developers to identify, reproduce and fix the corresponding bugs. 
    
    \item \textbf{Environmental issues (C2):} 
    The environmental issues are related to the problems of the network, CPU, memory, operating system, etc. 
    When the environmental issues occur, software testers are responsible to diagnose the system environment. 

    \item \textbf{Test script issues (C3):}
    The test script issues related to the defects within the test scripts (e.g., expressions, arguments, statements).
    When test script issues occur, software testers are responsible to diagnose and debug the test scripts. 
    \item \textbf{Third party library issues (C4):}
    The third-party library issues are associated with defects or incompatible problems in the third-party libraries, e.g., there are problems regarding the automatic logging system.
    When third-party library problems appear, it is necessary to ask developers to diagnose the third-party library software.
\end{itemize}

The distribution of the above four types of test failure causes are summarized in Table~\ref{tab:failure_causes}.
From the table, we can see that there is an unequal distribution of different failure causes among test logs.
For example, the vast majority of the failure causes are the \textbf{C1} (\textit{bug related issues}) and \textbf{C2} (\textit{environmental issues}), which make up a very large proportion (i.e., over 86\%) of all the failed test cases. 
The number of failed tests caused by \textbf{C3} (test script issues) and \textbf{C4} (third-party library issues) only account for a relatively small number of the failed test cases.
Especially regarding \textbf{C4}, only 101 test failures are caused by the \textit{third-party library issues}, which comprises only 1.4\% among all failed test cases. 

\begin{table}%[tbp]
\caption{Different Types of Failure Causes}
\label{tab:failure_causes}
% \vspace*{-10pt}
\begin{center}
\begin{tabular}{ccrr}
    \toprule
    {\bf ID} & {\bf Failure Causes} & {\bf Count} & {\bf Percentage} \\
    \midrule
    \textbf{C1}  & $\textbf{bug related issues}$ & $4,559$ & $63.7\%$ \\
    
    \textbf{C2}  & $\textbf{Environmental issues}$ & $1,664$ & $23.2\%$ \\
    
    \textbf{C3}  & $\textbf{Test script issues}$ & $835$ & $11.7\%$ \\
    
    \textbf{C4}  & $\textbf{Third party library issues}$ & $101$ & $1.4\%$ \\
    
    % \textbf{C4}  & $\textit{Tool problem}$ & $26$ & $50.9\%$ \\
    \midrule
    \textbf{Sum}  & $\textit{-}$ & $7,159$ & $100.0\%$ \\
    \bottomrule
\end{tabular}
% \vspace{-10pt}
\end{center}
\end{table}

\subsection{Key Insights}
Our approach has been inspired by the following four key insights, which lead to our solution for this task. 

\begin{figure}
% \vspace{-7pt}
\centerline{\includegraphics[width=0.50\textwidth]{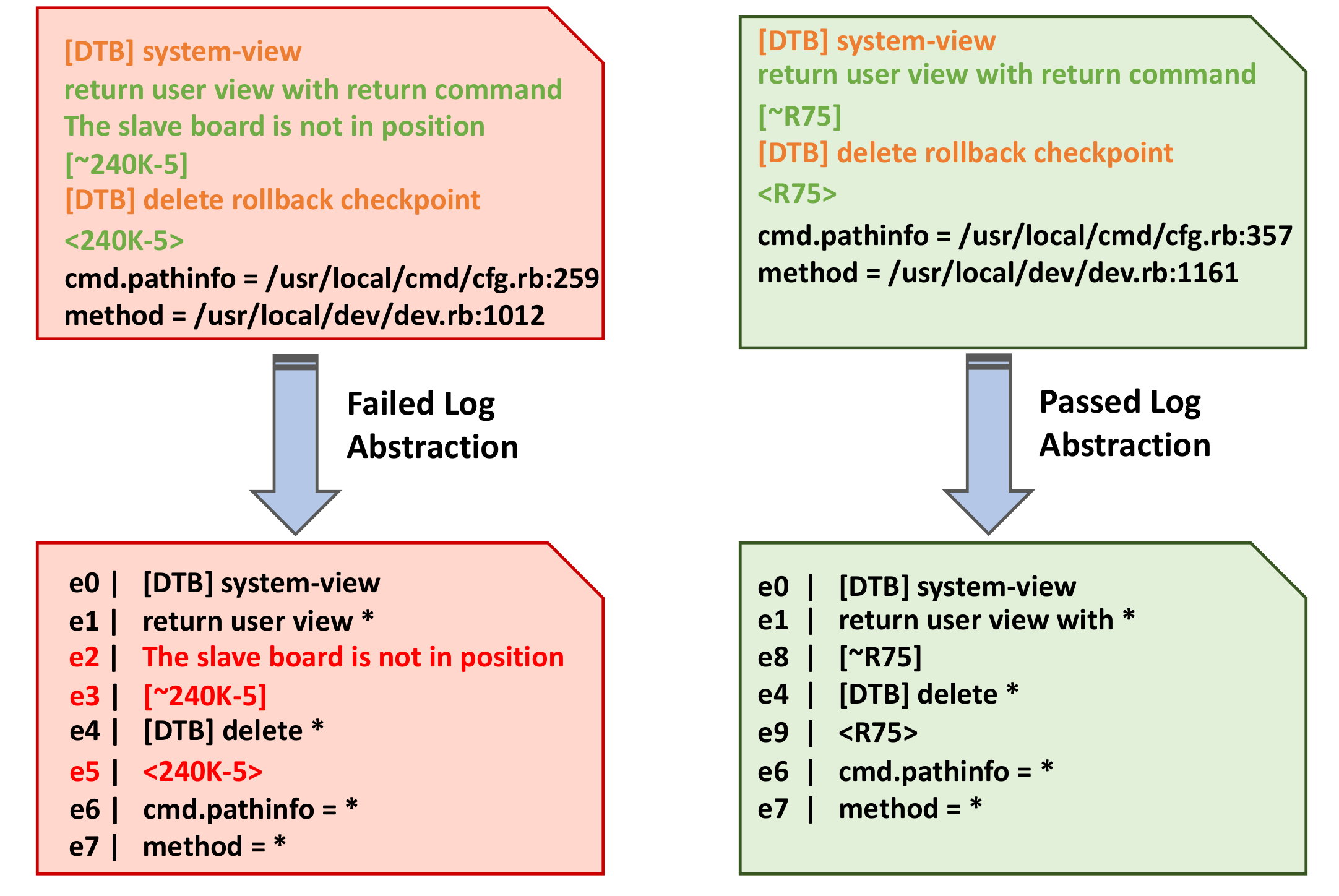}}
% \vspace*{-5pt}
\caption{Log abstraction of failed/passed logs}
% \vspace{-10pt}
\label{fig:log_abstraction}
\end{figure}

\noindent\textbf{Key Insight 1: Logs are often too large and too unstructured to analyze manually.}
First of all, the log files are often very large. 
For example, as shown in Table~\ref{tab:data_overview}, the failed test log contains 3,905 log lines on average, with the largest log file containing over 550K log lines.
Testers have to go through the entire log file to identify the log lines that correspond to the test failure. 
The sheer amount of log data makes its analysis a time-consuming and challenging task. 
Moreover, the log files are highly complex and unstructured. 
We present two snippets of failed/passed logs in Fig.~\ref{fig:log_abstraction}. 
As shown in Fig.~\ref{fig:log_abstraction}, a test log often involve a sequence of test steps (e.g., \textit{system-view} and \textit{delete rollback checkpoint} shown in orange text), each test step is followed by multiple lines of echo messages, which contains the state of the object (e.g., $\langle\textit{240K-5}\rangle$ and $\langle\textit{R75}\rangle$), environment variables (e.g., \textit{cmd.pathinfo}), exception messages (e.g., \textit{The slave board is not in position}), etc.

% The log abstraction reduces the specific lines that indicates a fault. 
% Log abstraction reduces the number of unique lines in a log. 
% Although the logs do not have a specific format, they contain static and dynamic parts. 
% The dynamic run specific information, such as date and test machine, can obscure the higher level patterns. 
% By removing this information, abstract lines contain the essence each line without any noisy details. 

To overcome this challenge, we adopt a log abstraction technique in our approach.
Log abstraction can significantly reduce the number of unique lines in a log. 
The raw log lines often contain dynamic run specific information (e.g., file path, IP address) that can hinder the automatic analysis, log abstraction can extract the log event by abstracting away dynamic parameters. 
For example, in Fig.~\ref{fig:log_abstraction}, the second last log line of the failed log (``\path{cmd.pathinfo=/usr/local/cmd/cfg.rb:259}'') and the passed log (``\path{cmd.pathinfo=/usr/local/cmd/cfg.rb:357}'') can be abstracted with the same log event $\mathbf{e_6}$ (\path{cmd.pathinfo = *}). 
After removing the superfluous information, the abstract log line only contains the essence of each log line without noisy details.
As a result, each log will be abstracted a sequence of log events which record the system status and operations. 
For example, the failed log snippet is abstracted into eight log events (i.e., $\mathbf{e_0}$ - $\mathbf{e_7}$) after the log abstraction process.

\noindent\textbf{Key Insight 2: 
The failed log events are more important than the passed log events for identifying failures causes. 
} 
We make the following observations: test failure locations should be contained in the lines of a failed log. 
On the contrary, a passed log should not contain the lines related to a failure. 
Therefore, log lines that appear in both failed test logs and passed test logs are unlikely to link to a fault. 
For a failed test log, most of the log lines record normal operations, only a small percentage of log lines are problematic and indicate problems. 
For example, as shown in Fig.~\ref{fig:log_abstraction}, the failed log contains a log event $\mathbf{e_1}$ (i.e., \texttt{return user view *}), 
However, the passed log also contains this log event, so it is unlikely that the test failure is related to the log event $\mathbf{e_1}$. 
In contrast, the log event $\mathbf{e_2}$ (\texttt{the slave board is not in position}) occurs only in the failed log, indicating the potential cause for this failure. 
Based on our assumption, we should focus on the log events that are only covered by failed test logs.

\noindent\textbf{Key Insight 3:
The single-problem log events are more important than the multi-problem log events for identifying failure causes. 
}
As mentioned above, a log event can occur in both failed test logs and passed test logs. 
Similarly, a log event can be relevant to a single failure cause or multiple failure causes. 
Based on the number of failure causes, we further divide the failed log event into single-problem log events and multi-problem log events. 
That is, if a log event is only relevant to one particular type of failure cause, we then consider this log event as a single-problem log event. 
Otherwise, if a log event is relevant to two or more types of failure causes, we consider this log event as a multi-problem log event. 
For example, as shown in Fig.~\ref{fig:workflow}, the log event $\mathbf{e_2}$ is associated with four types of failures (i.e., C1-C4), which is a multi-problem log event. 
The log event $\mathbf{e_3}$ is only relevant to C2 (\textit{environmental issues}) in history, which is a single-problem log event. 
In this study, we assume that multi-problem log events are less helpful in identifying the root cause of a specific test failure.
In contrast, the single-problem log event that only happens with respect to a particular failure, is more likely to be useful in failure causes prediction. 

\noindent\textbf{Key Insight 4: 
The minority-class log events are more important than the majority-class log events for identifying failure causes. 
}
Our last insight comes from the challenge of the imbalanced dataset. 
As discussed in Section~\ref{sec:pre}, each failed test log is manually labeled with a failure cause verified by testers. However, most of the failed logs are associated with the C1 (\textit{bug related issues}) and C2 (\textit{environmental issues}) (more than 86\%). Only a small percentage of failed logs are caused by C3 (\textit{test script issues}) and C4 (\textit{third party library issues}). 
In this study, regarding the single-problem log events, if the log event is associated with C1 or C2, we consider this log event as a majority-problem log event; if the log event is associated with C3 or C4, we consider this log event as a minority-problem log event. 
For example, as shown in Fig.~\ref{fig:workflow}, the log event $\mathbf{e_3}$ is a majority-problem log event (associated with C1 only), while the log event $\mathbf{e_8}$ is a minority-problem log event (associated with C4 only). 
When predicting the potential failures, even though $\mathbf{e_3}$ and $\mathbf{e_8}$ both occur twice in history, considering the C4 (\textit{third party library issues}) is comparatively rare than the C1 (\textit{bug related issues}), the minority-problem event log $\mathbf{e_8}$ clearly has greater predictive power for identifying failures causes.

\section{OUR APPROACH}
\label{sec:approach}
We first define the task of identifying test failure causes for our study. 
We then present the details of our proposed approach. 
The overall framework of our approach is illustrated in Fig.~\ref{fig:workflow}.

\subsection{Task Definition}
The goal of our work is to automatically predict the failure cause of a test based on analyzing the test logs. 
In particular, given the historical passed logs $\mathbf{P}$, the historical failed logs $\mathbf{F}$ and the associated failure causes $\mathbf{C}$, the failure cause prediction problem is to predict $\mathbf{C_{new}}$ for the newly arisen test failures via analyzing the unseen failed log $\mathbf{F_{new}}$ with the help of $\langle \mathbf{P}, \mathbf{F}, \mathbf{C} \rangle$. 
We formulate this task as a multi-class classification problem. 
More formally, our task is to find a function \texttt{Predict} so that:
\begin{equation}
\label{eq:patch}
\mathbf{Predict} ( \mathbf{F_{new}} |\langle \mathbf{P}, \mathbf{F}, \mathbf{C} \rangle) = \mathbf{C_{new}}
\end{equation}
% We formulate this task as a multiclass classification problem due to various causes for test failures. 
% The multiclass classification problem aims to classify instances into one out of more than two classes. 
% In this study, the newly failed test logs are instances for classifying, and their causes are target multiple classes. 

\subsection{Approach Details}
\begin{figure}
% \vspace{-7pt}
\centerline{\includegraphics[width=0.50\textwidth]{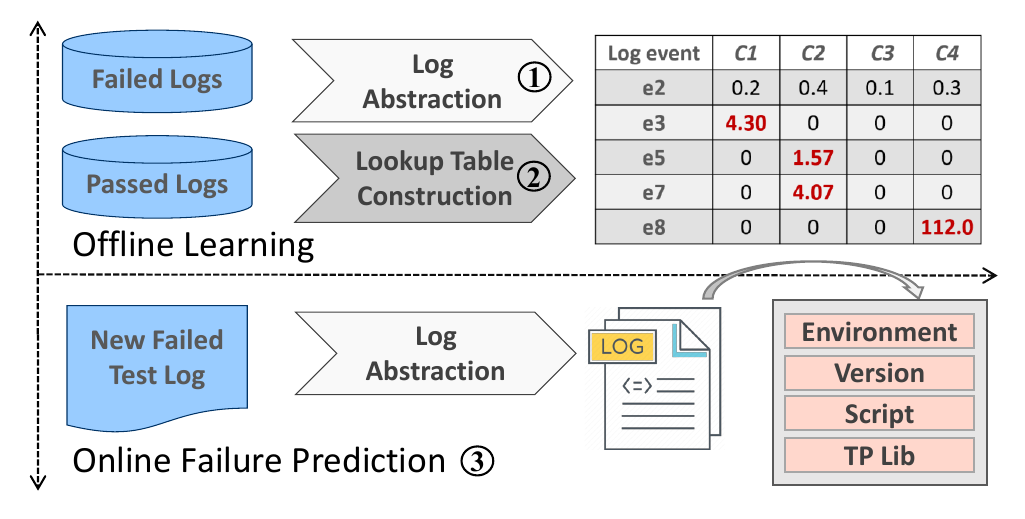}}
% \vspace*{-5pt}
\caption{Workflow of our approach}
% \vspace{-10pt}
\label{fig:workflow}
\end{figure}

In this paper, we propose {\sc NCChecker}, whose overall framework is presented in Fig.~\ref{fig:workflow}. 
{\sc NCChecker} consists of three stages: log abstraction, lookup table construction and failure causes prediction. 
In short, the unstructured raw logs are parsed into a sequence of log events by log abstraction. 
Then based on our previous insights and observations, we propose heuristic rules to make a lookup table. 
The lookup table records the predictive power scores for different log events regarding different failure causes. 
Finally, during the failure cause prediction stage, for a newly failed test log, {\sc NCChecker} can easily infer the root cause by checking from the lookup table. 
More details are presented in the following sub-sections. 

\subsubsection{Log Abstraction.}

\begin{figure*}
% \vspace{-7pt}
\centerline{\includegraphics[width=0.99\textwidth]{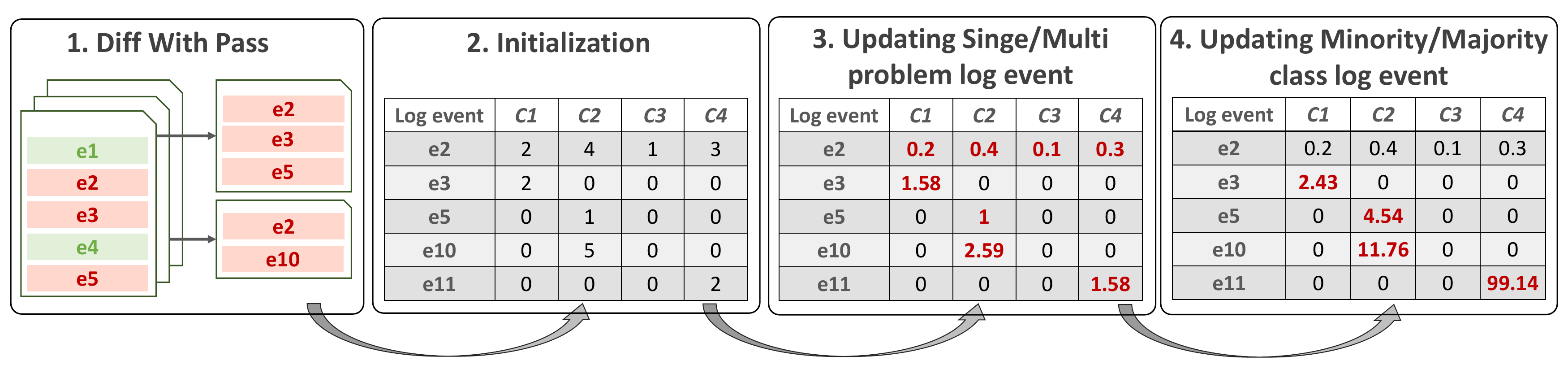}}
% \vspace*{-5pt}
\caption{Lookup Table Construction of {\sc NCChecker}} 
% \zp{To be Updated}} 
\label{fig:lookuptable}
% \vspace{-8pt}
\end{figure*}

As discussed as \textbf{Key Insight 1}, the raw logs are too large and unstructured to analyze manually. 
To overcome the unstructured nature of test logs and reduce the large amount of log messages, we first use log abstraction techniques to parse unstructured, free-text log messages into structured log events. 
The common way of log abstraction in industry is to write regular expression according to the logging statements in the source code. 
However, due to the rapidly growing log data and frequently software updates, it is too costly and time-consuming to manually construct regex.
Therefore, automatic log parsing without source code is necessary. 
Previous studies have proposed different log abstraction techniques~\cite{messaoudi2018search, makanju2011lightweight, hamooni2016logmine, du2016spell, dai2020logram}, in this study, we adopt the widely used log parser, Drain~\cite{he2017drain}, for our log abstraction tasks. 

In general, each log line contains two parts: the static part and the dynamic part. 
The static part describe the semantic meaning of a program event, while the dynamic part includes parameters (e.g., file name) that record system attributes. 
The goal of automatic log abstraction is to separate the static part from the dynamic part. 
In particular, the log parser can abstract a group of similar log messages into a unique log event, and mask the dynamic part with a placeholder (usually by an asterisk). 
For example, by applying the Drain log parser, the log message \textit{``Took 10 seconds to build instances''} can be abstracted with the log event ``\texttt{Took * seconds to build instances}'', where the static part represents the common part of similar log messages and the asterisk represents the dynamic part. 

After log abstraction, each log line is transformed into a unique log event, and a log file is transformed into a sequence of log events. 
As a result, for a given passed log $\textbf{p}$, it will be parsed into a sequence of log events, i.e., $\mathbf{E_{p}} = \{\mathbf{e_{p_{1}}}, \mathbf{e_{p_{2}}}, ..., \mathbf{e_{p_{m}}} \}$. 
Similarly, a failed log $\textbf{f}$ will be parsed into $\mathbf{E_{f}} = \{\mathbf{e_{f_{1}}}, \mathbf{e_{f_{2}}}, ..., \mathbf{e_{f_{n}}} \}$.
For example, as shown in Fig~\ref{fig:log_abstraction}, the failed log snippet is sequentially represented by its corresponding log events sequence $\{\mathbf{e_0}, \mathbf{e_1}, ..., \mathbf{e_7}\}$.  

\subsubsection{Lookup Table Constructing.}
After parsing all logs (including passed logs and failed logs) into log events, in this step, we construct a lookup table for representing the predictive power of each log event with respect to different failure causes. 
The lookup table is constructed by the following steps: 

\textbf{1. Diff with pass.}
Log events that both occur in passed log and failed log are unlikely to reveal a fault, which may introduce noise for failure causes prediction. 
Inspired by our \textbf{Key Insight 2}, we assume that the failed log events are more important than the passed log events for identifying failure causes.  
After log abstraction, each passed log is parsed into a set of log events, we collect all log events occur in passed logs to construct a passed log events pool, $\mathcal{P}$. 
Similarly, we collect all the log events occur in failed logs to construct a failed log events pool, $\mathcal{F}$. 
In this step, we perform \textit{Diffwithpass} operation to remove all log events that appear in passed logs from failed logs. 
Specifically, for each log event in $\mathcal{F}$, if it also appears in $\mathcal{P}$, we then remove this log event from $\mathcal{F}$. 
After performing the \textit{Diffwithpass} operation, all the log events in $\mathcal{F}$ occur only in failed logs, denoted as $\mathcal{F'}$. 
For example, as shown in Fig.~\ref{fig:log_abstraction}, only the log events $\mathbf{e_2}$, $\mathbf{e_3}$, $\mathbf{e_5}$ (highlighted in red color) are retained for subsequent log analysis, while other log events (e.g., $\mathbf{e_0}$, $\mathbf{e_1}$) are removed. 

\textbf{2. Lookup table initialization.}
In this step, for each log event in $\mathcal{F'}$, we count its failed times with respect to different failure causes. 
Specifically, we initialize a lookup table $\mathbb{T}$, where the rows are log events in $\mathcal{F'}$, the columns are different failure causes $\mathbf{C_{i}} (1 \leq i \leq 4)$. 
The cell value $c_{ij}$ represents the number of times that the log event $e_{i}$ failed according to the failure cause $\mathbf{C_{j}}$.

In particular, the lookup table $\mathbb{T}$ initialization is conducted as follows: 
First, all the cell values in table $\mathbb{T}$ are initialized to zero.  
Given the failed test log $\mathbf{f_{i}}$ and its failure cause $\mathbf{C_{i}}$, for each log event $\mathbf{e_{k}}$ in $\mathcal{F'}$, if $\mathbf{e_{k}}$ occurs in $\mathbf{f_{i}}$, we then increment the count by 1 regarding the log event $\mathbf{e_{k}}$ and the failure cause $\mathbf{C_{i}}$.
After the table initialization, the lookup table $\mathbb{T}$ records the failed frequency of all log events in $\mathcal{F'}$ regarding different failure causes. 
For example, as shown in Fig.~\ref{fig:lookuptable}, the log event $\mathbf{e_2}$ occurs 10 times in total, two of which are associated with \textit{bug related issues} (C1), four times with \textit{environmental issues} (C2), once with \textit{test script issues} (C3) and three times with \textit{third party library issues} (C4). 
The log event $\mathbf{e_{10}}$ occurs five times in total, all of which are associated with \textit{environmental issues} (C2).

\textbf{3. Lookup table updating with single/multi-problem log events.} 
Inspired by our \textbf{Key Insight 3}, we assume the single-problem log events are more important than the multi-problem log events. 
Therefore in this step, we would like to increase the predictive power of single-problem log events and decrease the predictive power of multi-problem log events. 
To do this, we first normalize the multi-problem log event frequency values to the range of $(0, 1)$. 
In particular, regarding the multi-problem log event $\mathbf{e_{i}}$, the updated cell value $c_{ij}'$ is calculated as follows: 

\begin{equation}
c_{ij}' = \frac{c_{ij}}{\sum^{4}_{j=0}c_{ij}}
\end{equation}

For example, the log event $\mathbf{e_2}$ is associated with several failure causes, which is a multi-problem log event. 
After updating the multi-problem log event, the original count value of $e_2$ (i.e., $[2, 4, 1, 3]$) will be normalized as 
$[0.2, 0.4, 0.1, 0.3]$, which represents the 20\% of $e_2$ contributes to C1.

In contrast, regarding the single-problem log event $\mathbf{e_{k}}$, the updated cell value $c_{kj}'$ is calculated as follows: 
\begin{equation}
c_{kj}' = \left\{ \begin{array}{lcr} 
                0.0 & \mbox{if} & c_{kj} = 0 \\ 
                1.0 & \mbox{if} & c_{kj} = 1 \\ 
                log_{2}{(1 + c_{kj})} & \mbox{if} & c_{kj} > 1
\end{array}\right.
\end{equation}
For example, the log event $\mathbf{e_3}$ and $\mathbf{e_5}$ are only associated with C2, which are single-problem log events.
The original count value will be updated. 
% The log event YYY only failed once regarding the $C_{2}$, 
% For example, the log event XXX is associated with several failure causes, which is a multi-problem log event. 
% After updating, the original frequency value $[3, 5, 4, 1]$ will be normalized as $[0.25, 0.25, 0.4, 0.1]$, which represents 25\% of this log event is associated with $C_{1}$, 30\% and 30\% are associated with $C_{2}$ and $C_{3}$ respectively. 

% Regarding the single-problem log events, we firstly set its corresponding failure cause value to 1.0.
% The original frequency value 
% which represents 100\% of this log event is associated with failure cause $C_{2}$. 
% To further distinguish the predictive power of log event YYY and log event ZZZ, we 

\textbf{4. Lookup table updating with minority/majority-class log events.} 
Inspired by our \textbf{key insight 4}, we assume the minority-class log events are more important than the majority-class log events. 
Therefore in this step, we aim to increase the predictive power of the minority-class log events, and decrease the predictive power of the majority-class log events. 
To do this, we give different weights to different log events for different log events based on whether they belong to the majority or the minority classes.
In particular, we use the ICF (inverse class frequency) to operationalize the importance of a log event to different class frequencies. 
Regarding a single log event $\mathbf{e_{i}}$, the updated cell value $c_{ij}''$ is calculated as follows:
\begin{equation}
c_{ij}'' = c_{ij}' \times ICF_{\mathbf{C_j}}
\end{equation}

\begin{equation}
ICF_{\mathbf{C_j}} = \frac{N}{N_{\mathbf{C_j}}}
\end{equation}
where $ICF_{\mathbf{C_j}}$ represents the inverse class frequency value with respect to failure cause $\mathbf{C_{j}}$. 
$N$ denotes the total number of failed test logs and $N_{\mathbf{C_{j}}}$ denotes the number of failed test logs that belong to $\mathbf{C_{j}}$. 
The intuition here is to assign a higher weight to the log events associated with minor classes. 
For example, the log event $\mathbf{e_{11}}$ is associated with failure cause $\mathbf{C_{4}}$, which is a minority-class log event. Therefore its value will be updated by multiplying $ICF_{\mathbf{C_4}}$ (i.e., 62.75). For the majority-class log event $\mathbf{e_3}$, its value will be updated by multiplying $ICF_{\mathbf{C_{1}}}$ (i.e., 1.54). 
After updating, the log event $\mathbf{e_{11}}$ has a greater predictive power than $\mathbf{e_3}$ due to rare class frequency. 

% all train log events: 112099
% all test log events: 17912
% version weights: 1.54
% env weights: 4.54
% script weights: 8.56
% lib weights: 62.75

After the initialization and updating process, the lookup table $\mathbb{T}$ represents the predictive power of different log events for different failure causes. 
The higher the value, the stronger the indicator the log event is associated with the failure cause. 

\subsubsection{Failure cause prediction.}
After constructing the lookup table $\mathbb{T}$, we can easily check out the matching score between a failed log event and a specific failure cause. When it comes to the failure cause prediction stage, for a given newly failed test log, we first perform the log abstraction to parse the log into a sequence of log events. 
After that, for each log event in the sequence, if the log event appears in the lookup table $\mathbb{T}$, we check out the matching score of this log event with respect to different failure causes. 
Finally, we sum up all the associated row values, the failure cause with the highest score will be reported as the final prediction result.
The log events with the maximum values will be highlighted for investigation.

\section{Empirical Evaluation}
\label{sec:eval}
In this section, we first present the baselines, the evaluation metrics and our experiment settings. 
We then describe the results of our automatic evaluation results. 
In this study, we aim to answer the following four research questions: \begin{itemize}
    \item RQ-1. How effective is our approach for the failure causes prediction task?
    \item RQ-2. How effective is our approach for predicting different types of failure causes?
    \item RQ-3. How effective is our use of heuristic rules for constructing lookup tables?
    \item RQ-4. How effective is our approach in terms of using computation resources?
\end{itemize}
% Then, we describe , research questions (RQs), and the corresponding experimental results.
% \subsection{Research Questions}

\subsection{Baselines} 
To demonstrate the effectiveness of our proposed model, {\sc NNChecker}, we compared it with the following chosen baselines:
\begin{itemize}
    
    \item \textbf{Random Guess (RG).}
    % RG is usually adopted as a baseline when there is no previous method for addressing the same research question. 
    Regarding the test failure cause prediction, the RG model randomly determine a failure cause for a given failed test log according to the data distribution. 
    In particular, we predict the failure cause randomly and we repeat the random prediction 100 times to get a median performance. 
    
    \item \textbf{The Majority Class Classifier (MCC).}
    Considering the data imbalance problem in our dataset, e.g., 63\% of the test failures are caused by bug related issues, it is reasonable to use the majority class classifier as a baseline for comparison purposes. 
    In particular, the majority classifier model predicts the most frequent class label (i.e., bug related problem in our study) for all test samples.

    \item \textbf{CAM.}
    Jiang et al.~\cite{jiang2017causes} proposed a a novel approach CAM (Cause Analysis Model) that infers test failure causes by analyzing test logs. 
    CAM runs TF-IDF across the logs to determine which terms had the highest importance, it then construct attribute vectors based on test log terms. 
    When a new test alarm occurs, CAM calculates the log similarity between the new test log and the historical test logs. 
    Finally, the unseen failed test log is categorized by examining the categories of the K nearest neighbours.

    \item \textbf{LogFaultFlagger.}
    Amar et al.~\cite{amar2019mining} recently proposed an approach named LogFaultFlagger to predict bugs and localize faults in test logs via mining historical test logs. 
    LogFaultFlagger % modifies the TF-IDF to identify the most relevant log lines related to past product failures. 
    vectorizes the logs with line-IDF metrics and uses EKNN to identify which logs are likely to lead to product faults and which lines are the most probable indication of the failure. 
    We adapt the LogFaultFlagger for our task of failure cause prediction. 
    % To be more specific, for a given test log, four different possible vectors with respect to  are constructed, 
\end{itemize}

\subsection{Evaluation Metrics}
Since our task is a multi-class classification problem, we adopted the widely-accepted evaluation metrics, i.e., Precision, Recall and F1-score to evaluate the performance of {\sc NCChecker} and baseline methods. 
We define the following statistics with respect to our task:
(i) $TP_{k}$ (True Positives): $TP_{k}$ is the number of test logs assigned correctly to class $k$. 
(ii) $FP_{k}$ (False Positives): $FP_{k}$ is the number of test logs that do not belong to class $k$ but assigned to class $k$ incorrectly by classifier. 
(iii) $FN_{k}$ (False Negatives): $FN_{k}$ is the number of test logs that do not assigned to class $k$ by classifier but which actually belong to class $k$. 
Our evaluation metrics are defined as follows: 
\begin{itemize}

    \item \textbf{Precision:} 
    Precision is the fraction of true positive samples divided by the total number of positively predicted samples (column sum). 
    The precision metric for class $k$, namely $p_{k}$, is defined as follows:
    
    \begin{equation}
    p_{k} = \frac{TP_{k}}{TP_{k} + FP_{k}}
    \end{equation}
    
    The precision over all $K$ categories (macro average precision) is defined as follows: 
    
    \begin{equation}
    Precision = \frac{\sum^{K}_{k=1} p_{k}}{K}
    \end{equation}
    
    \item \textbf{Recall:} 
    Recall is the fraction of true positive samples divided by the total number of positively classified samples (row sum). 
    Similarly, the recall metric for class $k$, namely $r_{k}$, is defined as follows:  
    % Recall represents the proportion of all resolved task comments that are correctly classified as resolved. The Recall metric is defined as follows: 
    
    \begin{equation}
    r_{k} = \frac{TP_{k}}{TP_{k} + FN_{k}}
    \end{equation}
    
    The recall over all categories (macro average recall) is defined as follows: 
    
    \begin{equation}
    Recall = \frac{\sum^{K}_{k=1} r_{k}}{K}
    \end{equation}

    \item \textbf{F1-score:} F1-score is the harmonic mean of precision and recall, which  combines both of the two metrics above. 
    It evaluates if an increase in precision (or recall) outweighs a reduction in recall (or precision), respectively. 
    The F1-score metric for class k, namely $f1_{k}$, is defined as follows: 
    
    \begin{equation}
    f1_{k} = \frac{2 \times p_{k} \times r_{k}} {p_{k} + r_{k}}
    \end{equation}
    
    The F1-score over all categories (macro average f1-score) is defined as follows: 
        
    \begin{equation}
    F1 = \frac{\sum^{K}_{k=1} f1_{k}}{K}
    \end{equation}
    
\end{itemize}

The higher an evaluation metric, the better a method performs. Note that there is a trade off between Precision and Recall. F1-score provides a balanced view of precision and recall. 
It is worth mentioning that we did not consider accuracy due to data imbalance.

\subsection{Experimental Settings}
We divide our dataset into training set and testing set, the training set is used for learning and building the lookup table, and the testing set is hold out to evaluate the performance of our approach (we did not use validation in this study since there are no hyper parameters of our model for fine tuning).
In particular, we randomly sampled 10\% of failed test logs for testing and kept the rest for training. 
The details of the training and testing samples as well as the log events are summarized in Table~\ref{tab:data_sta}. 
All experimental results are conducted over a server equipped with Intel(R) Core(R) CPU i7-4790 at 3.60GHZ on 32GB RAM, four cores, and 64-bit Windows 10 operating system. 

\begin{table}%[t]
\caption{Statistics of our training/testing Datasets }
\label{tab:data_sta}
%\vspace*{-10pt}
\begin{center}
% {\scriptsize
% \revv{
\begin{tabular}{lcr}
    \toprule
    {\bf Type} & {\bf Measurement} & {\bf Count}  \\
    \midrule
    \multirow{4}{*}{\bf C1. Bug related}
    & Train Logs           & 2,600   \\ 
    & Test Log             & 289   \\ 
    & Train Log events     & 91,237   \\
    & Test Log events      & 11,916   \\ \midrule
    \multirow{4}{*}{\bf C2. Environment}
    & Train Logs           & 885   \\ 
    & Test Log             & 99   \\
    & Train Log events     & 26,388   \\ 
    & Test Log events      & 5,112   \\ \midrule
    \multirow{4}{*}{\bf C2. Test Script}
    & Train Logs           & 470   \\ 
    & Test Log             & 47   \\ 
    & Train Log events     & 18,724   \\ 
    & Test Log events      & 5,062   \\ \midrule
    \multirow{4}{*}{\bf C2. TP Lib}
    & Train Logs           & 65   \\ 
    & Test Log             & 8   \\ 
    & Train Log events     & 5,166   \\ 
    & Test Log events      & 1,563   \\ \bottomrule
\end{tabular}
% } % }
\end{center}
\end{table}

% We implemented {\sc TDCleaner} in Python using the Pytorch framework. 
% We used the pre-trained BERT model~\cite{devlin2018bert} as the encoder for embedding training samples, which provides a powerful context-dependent sentence representation. 
% BERT can be easily extended to a joint classification model.
% In our model, \textit{Code Change Encoder}, \textit{TODO Comment Encoder} and the \textit{Commit Message Encoder} are jointly trained to minimize the cross entropy. 
% After the encoding process, \textit{code\_change}, \textit{todo\_comment} and \textit{commit\_msg} will be mapped to a 768 dimensional vector respectively. 
% During the training phase, we optimized the parameters of our model (including the BERT parameters and MLP parameters) using Adam~\cite{kingma2014adam} with a batch size of 32.
% We use the ReLu as the activation function and employ three hidden layers for MLP. 
% A dropout~\cite{srivastava2014dropout} of 0.2 is used for dense layers before computing the final probability. 
% The model is validated every 1,000 batches on the validation set with a batch size of 32. 
% We set the learning rate of Adam to 0.001 and clip the gradients norm by 2. 
% The model with the best performance on the validation set was used for our evaluations. 
% The learning rate is decayed by a 

\subsection{Quantitative Analysis}
\label{subsec:quantitative_eval}
\subsubsection{RQ1: The Effectiveness Evaluation}
To evaluate the effectiveness of our proposed model, i.e., {\sc NCChecker}, we compare {\sc NCChecker} with the baseline methods on our testing set in terms of overall Precision, Recall and F1-score.
The evaluation results is shown in Table~\ref{tab:effective_eval}. 
From the table, we have the following observations:
\begin{itemize}

    \item The currently state-of-the-art models, e.g., CAM and LFF performs suboptimal on our evaluation set.
    The CAM creates TF-IDF vectors based on the raw log messages and ranks the logs using cosine similarity. 
    While the LFF parses the raw logs into log events and creates the TF-Line IDF vectors based on the parsed log events. 
    Considering the constructed TF-IDF vectors are relatively sparse, the retrieval-based models (i.e., CAM and LFF) can easily ignore the critical log events, which contain the key information for indicating the failure root cause correctly.

    \item Our model, i.e., {\sc NCChecker}, outperforms all the baseline methods by a large margin in terms of all evaluation metrics. 
    We attribute this to the following reasons: First, it uses the log abstraction techniques to parse the unstructured log lines into structured log events, instead of recording the overwhelming amount of log messages, {\sc NCChecker} transforms a log file into a sequence of log events, which can extract useful patterns from raw log messages. 
    Second, compared with the information retrieval based models, which rely on the test log from existing database, our model predicts the failure causes by using finer-granularity log event elements. 
    We designed heuristic rules to boost the predictive power of the critical log events, when two raw logs are dissimilar but share a number of critical log events, our model can easily make the correct predictions by checking these critical log events. 
    
\end{itemize}

\begin{table}%[tbp]
\caption{Overall Effectiveness Evaluation}
\label{tab:effective_eval}
% \vspace*{-10pt}
\begin{center}
\begin{tabular}{crrr}
    \toprule
    {\bf Measure} & {\bf Precision} & {\bf Recall} & {\bf F1} \\
    \midrule
    \textbf{RG}  & $25.3\%$ & $25.0\%$ & $19.7\%$ \\
    
    \textbf{MCC}  & $25.0\%$ & $16.4\%$ & $19.8\%$ \\
    
    \textbf{CAM}  & $53.1\%$ & $61.8\%$ & $55.0\%$ \\
    
    \textbf{LFF}  & $37.2\%$ & $48.3\%$ & $37.8\%$ \\
    
    {\sc \textbf{NCChecker}} & $\textbf{72.0\%}$ & $\mathbf{72.4\%}$ & $\mathbf{71.9\%}$ \\
    \bottomrule
\end{tabular}
% \vspace{-0.5cm}
\end{center}
\end{table}

% \begin{table}%[tbp]
% \caption{Effectiveness Evaluation (Java)}
% \label{tab:effective_eval_java}
% % \vspace*{-10pt}
% \begin{center}
% \begin{tabular}{|c|c|c|c|c|}
%     \hline
%     {\bf Measure} & {\bf Accuracy} & {\bf Precision} & {\bf Recall} & {\bf F1} \\
%     \hline\hline
%     \textbf{TCO}  & $56.5\%$ & $47.5\%$ & $58.7\%$ & $52.5\%$ \\
%     \hline
%     \textbf{TMO}  & $56.9\%$ & $23.4\%$ & $73.3\%$ & $35.5\%$ \\
%     \hline
%     \textbf{TCMO}  & $59.7\%$ & $57.6\%$ & $60.8\%$ & $59.2\%$ \\
%     \hline
%     \textbf{IRSC}  & $60.1\%$ & $59.0\%$ & $70.1\%$ & $64.0\%$ \\
%     \hline
%     {\sc \textbf{TDCleaner}}  & $\mathbf{85.0\%}$ & $\textbf{86.2\%}$ & $\mathbf{84.4\%}$ & $\mathbf{85.3\%}$ \\
%     \hline
% \end{tabular}
% % \vspace{-0.5cm}
% \end{center}
% \end{table}

\vspace{5pt}
\noindent
\framebox{\parbox{\dimexpr\linewidth-2\fboxsep-2\fboxrule}{
\textbf{Answer to RQ-1: 
How effective is our approach for test failure causes prediction? -- 
We conclude that our approach is highly effective for identifying the root causes of failed test logs.}}} 
\vspace{5pt}

\subsubsection{RQ2: Class-Wise Evaluation} 
The test failures are caused by a various of failure causes. 
Accurately identifying the failure causes are important because different actions need to be taken subsequently. 
One of the key challenges with respect to our task is that the class imbalance problem, i.e., the majority classes are more frequent occurring than the minority classes. 
For example, the \textbf{C1} (bug related issues) accounts for more than 63\% of all failed test logs. 
To verify the effectiveness of our model for identifying failure causes with respect to different failure cause classes, we conduct a class-wise evaluation in this research question. 
In particular, we evaluate the performance of {\sc NCChecker} and baselines regarding different types of failure causes. 
The evaluation results are shown in Table~\ref{tab:class_eval}. 
It can be seen that:

\begin{itemize}
    
    \item It is obvious that all the approaches (excluding ours) achieve a better performance on majority classes (e.g., \textbf{C1} and \textbf{C2}) than the minority classes (e.g., \textbf{C3} and \textbf{C4}). 
    This is because the majority class have more examples, the models can learn meaningful patterns from these abundant training samples, while it is more difficult to explore the minority class patterns given its smaller sample size and skewed class distribution.

    \item The RG and MCC model achieve the worst performance with respect to the minority class failure causes. 
    This is reasonable because the prediction results of RG and MCC model simply rely on the proportions of the majority classes. The likelihood of assigning a test failure into the minority classes is very small. 
    
    \item The CAM and LFF model have their advantage as compared to the RF and MCC model. 
    It is notable that CAM can achieve a comparable (i.e., \textbf{C1}) or better (i.e., \textbf{C2}) performance than our approach, 
    but its predictive performance on minority classes are still suboptimal. 
    This is because that the CAM and LFF  are information retrieval based models. Their performance heavily rely on whether similar test logs can be found and how similar the test logs are. 
    Considering the limited number of test logs belong to the minority classes, similar logs are hard to find in the training set. 
    
    \item 
    Regarding the minority class, it is obvious that our model, {\sc NCChecker}, outperforms all other baselines. 
    Rather than checking the similar test logs in history, {\sc NCChecker} maintains a lookup table by estimating the predictive power of the historical failed log events, we designed heuristic rules to increase the predictive power of the minority-class log events. 
    As a result, if the minority-class log events appear in the minority-class test logs, our model are more likely to infer correct failure causes regardless of the size of the minority-classes. 
    
    \item 
    There is still a large room for our approach to improve regarding minority-classes. 
    For example, the precision of our model for \textbf{C3} is 54.3\%, which means around half of the test script failure test logs are wrongly assigned. 
    The reason may be that there are log events only available in the testing set and not in the training set, our model is unable to handle the out-of-table log events. 

\end{itemize}

\begin{table}%[t]
\caption{Class-wise Effectiveness Evaluation}
\label{tab:class_eval}
%\vspace*{-10pt}
\begin{center}
% {\scriptsize
% \revv{
\begin{tabular}{lcrrr}
    \toprule
    {\bf Type} & {\bf Approach} & {\bf Precision} & {\bf Recall} & {\bf F1} \\
    \midrule
    \multirow{5}{*}{\bf C1. Bug related}
    & RG            & 26.7\% & 65.2\% & 37.9\% \\ 
    & MCC           & 65.6\% & \textbf{100.0\%} & 79.2\% \\ 
    & CAM           & 83.3\% & 81.4\% & 82.3\% \\ 
    & LFF           & 80.7\% & 71.9\% & 76.0\% \\ 
    & NCChecker     & \textbf{85.4\%} & 80.9\% & \textbf{83.1\%} \\ \midrule
    \multirow{5}{*}{\bf C2. Environment} 
    & RG            & 26.5\% & 23.4\% & 24.9\% \\ 
    & MCC           & 0.0\% & 0.0\% & 0.0\% \\ 
    & CAM           & \textbf{69.4\%} & \textbf{73.9\%} & \textbf{71.5\%} \\ 
    & LFF           & 35.3\% & 41.1\% & 38.0\% \\ 
    & NCChecker     & 60.2\% & 67.0\% & 63.4\% \\ \midrule
    \multirow{5}{*}{\bf C3. Test Script}   
    & RG            & 19.6\% & 9.4\% & 12.7\% \\ 
    & MCC           & 0.0\% & 0.0\% & 0.0\% \\ 
    & CAM           & 45.6\% & 42.0\% & 43.7\% \\ 
    & LFF           & 28.6\% & 30.3\% & 29.4\% \\ 
    & NCChecker     & \textbf{56.5\%} & \textbf{66.7\%} & \textbf{61.2\%} \\ \midrule
    \multirow{5}{*}{\bf C4. TP Lib}   
    & RG            & 2.9\% & 1.8\% & 3.3\% \\ 
    & MCC           & 0.0\% & 0.0\% & 0.0\% \\ 
    & CAM           & 14.3\% & 50.0\% & 22.2\% \\ 
    & LFF           & 4.1\% & 50.0\% & 7.7\% \\ 
    & NCChecker     & \textbf{85.7\%} & \textbf{74.9\%} & \textbf{80.0\%} \\ \bottomrule
\end{tabular}
% } % }
\end{center}
\end{table}

\vspace{5pt}
\noindent
\framebox{\parbox{\dimexpr\linewidth-2\fboxsep-2\fboxrule}{
\textbf{Answer to RQ-2: 
How effective is our approach for predicting different types of failure causes? -- 
We conclude that our approach is effective for failure causes prediction with respect to different failure types and can successfully handle the minority classes.}}} 
\vspace{5pt}

\subsubsection{RQ3: Ablation Evaluation} 
The key to our test failure cause prediction task is how effectively the lookup table can capture the relationship between different log events and failure causes. 
% To build the lookup table, we propose three heuristic rules to manage the predictive power of different log events. 
We propose three key heuristic rules guiding us to build the lookup table.
Specifically, we construct the lookup table by following four key steps: step1 (diff with pass), step2 (lookup table initialization), step3 (lookup table updating with single/multi-problem log events), step4 (lookup table updating with minority/majority-class log events).
To study the effectiveness of our three heuristic rules, we conduct an ablation analysis to evaluate their effectiveness and contributions one by one. 
We compare {\sc NCChecker} with three of its incomplete versions: 
\begin{itemize}
    \item \textbf{Drop 1:} 
    We drop step1 (diff with pass) in this version. 
    The lookup table is constructed by step2 (lookup table initialization), step3 (lookup table updating with single/multi-problem log events) and step4 (lookup table updating with minority/majority-class) only.  
    \item \textbf{Drop 2:} 
    In this version, we drop step3 (lookup table updating with single/multi-problem log events). 
    The lookup table is constructed by step1 (diff with pass), step2 (lookup table initialization), and step4 (lookup table updating with minority/majority-class) only.  
    \item \textbf{Drop 3:}
    In this version, we drop step4 (lookup table updating with minority/majority-class) and construct the lookup table only with step1 (diff with pass), step2 (lookup table initialization), step3 (lookup table updating with single/multi-problem log events). 
    \item {\sc \textbf{NCChecker:}}
    Our model which considers all the steps to construct the lookup table.
\end{itemize}

\begin{table}%[tbp]
\caption{Ablation Evaluation}
\label{tab:ablation_eval}
\vspace*{-10pt}
\begin{center}
\begin{tabular}{lrrr}
    \toprule
    {\bf Measure} & {\bf Precision} & {\bf Recall} & {\bf F1} \\
    \midrule
    \textbf{Drop 1} & $39.7\%$ & $58.3\%$ & $39.1\%$ \\
    \textbf{Drop 2} & $44.9\%$ & $36.7\%$ & $17.2\%$ \\
    \textbf{Drop 3} & $48.2\%$ & $78.1\%$ & $55.3\%$ \\
    {\sc \textbf{NCChecker}}  & $72.0\%$ & $72.4\%$ & $71.9\%$ \\
    \bottomrule
\end{tabular}
% \vspace{-30pt}
\end{center}
\end{table}

The experimental results are shown in Table~\ref{tab:ablation_eval}. 
From the table, several points stand out: 
\begin{itemize}
    \item \textbf{No matter which step we remove, the overall performance of our model decreases.}
    This shows the importance and usefulness of our key insights. All three assumptions provide valuable information to construct the lookup table respectively.
    \item \textbf{Drop 2 achieves the worst performance.} 
    It is clear that there is a significant drop overall in every evaluation measure after removing step3.
    This signals that the step3 (i.e., lookup table updating with single/multi problem log events) is the most important of all the steps for constructing the lookup table and has major contributions to the overall performance. 
    
%   \item \textbf{The performance of Drop 3 is better than the other two variants.} In other words, keeping step1 and step?? achieve a minimal performance drop. This justifies the importance and necessity of the above two steps in making the lookup table. 
\end{itemize}

\vspace{5pt}
\noindent
\framebox{\parbox{\dimexpr\linewidth-2\fboxsep-2\fboxrule}{
\textbf{Answer to RQ-3: 
How effective is our use of three heuristic rules for constructing lookup table? -- 
We conclude that all the three heuristic rules are effective and helpful to enhance the performance of our model.}}} 
\vspace{5pt}

\subsubsection{RQ4: Computation Resources Evaluation} 
Due to the huge number of test logs, the rapidly increasing computation resources (e.g., computation time and memory usage) are key challenges for model design. 
In this research question, we analysis the time consumption as well as memory consumption of our model. 

Regarding the time consumption, we record the training time and testing time of {\sc NCChecker}. 
The time consumption of {\sc NCChecker} on training is mostly for the log abstraction. 
It takes three hours to parse all the test logs (including failed test logs and passed test logs) using Drain. 
However, we argue that the log abstraction process is a one-time cost. 
The subsequent operations of making the lookup table is highly efficient. According to the Equations defined in Section~\ref{sec:approach}, the lookup table can be constructed by going through all the log events only once, which is dependent on the sizes of the training logs and generated log events. 

The time cost advantage of {\sc NCChecker} is more obvious in terms of the testing procedures.
Regarding the evaluation, the time complexity of IR based models are $O(N)$, while the time complexity of our model is $O(1)$. 
This is because for a given test log, IR based models calculate the similarity score across all training logs. 
For our approach, the failed reason can be predicted by checking out the associated log events scores from the lookup table directly. 
{\sc NCChecker} takes 8.75 seconds for analyzing the 443 test logs, which means it costs only 20ms on average to check each log.

Regarding the memory consumption, the raw log files are often very large. 
In particular, it cost over 3GB to store the raw log files.
After log parsing, all the log events and their frequency are recorded and the memory consumption is significantly dropped (from 3GB to 6M). 
By employing our approach, we only need manage and maintain a relatively small size lookup table (i.e., 8.9KB), which records all relevant scores between different log events and failure causes, for diagnosing test failures.

\vspace{5pt}
\noindent
\framebox{\parbox{\dimexpr\linewidth-2\fboxsep-2\fboxrule}{
\textbf{Answer to RQ-4: How effective is our approach in terms of the computation resources? -- 
We conclude that our approach is efficient and memory saving.}}} 
\vspace{5pt}
% 可解释性

\section{Related Work}
\label{sec:related}
In this section, we present the prior research that is related to this paper.

\subsection{Log-based Root Cause Analysis}
Root cause analysis, also known as failure diagnosis, aims to identify the underlying causes leading to a test failure that has affected end users. 
Root cause analysis is a crucial step for effectively resolving the software problems, which is extremely expensive and inefficient~\cite{he2021survey, zhang2019inflection, yu2014comprehending}. 
% Previous studies~\ref{} suggested that developers spend more than half of their time root cause analysis takes over 

As modern software system grows rapidly and becomes more mature, test failures are more and more difficult to analysis and diagnose~\cite{kavulya2012failure, chuah2010diagnosing, zhou2019latent}. 
For example, Jiang et al.~\cite{jiang2009understanding} reported that problem debugging is time-consuming and challenging which can be improved by using logs. 
They suggested developers to automate failure diagnosis process to speed up the problem fixing time. 
Zhou et al.~\cite{zhou2018fault} studied the failure debugging process with respect to microservice systems. 
They concluded that proper tracing and visualization techniques can improve failure diagnosis, which shows the necessity for intelligent log analysis tools. 

Researchers have developed different techniques for automating the log-based failure cause diagnosis~\cite{jiang2017causes, shang2013assisting, amar2019mining}. 
The works most similar to ours are the retrieval-based root cause analysis methods. 
In particular, retrieval-based methods retrieve similar failures in history for better diagnosing newly-occurred failures. 
Shang et al.~\cite{shang2013assisting} focused on diagnosing applications in Hadoop system by injecting failures manually and analyzing the logs. 
Nagaraj et al.~\cite{nagaraj2012structured} investigated the system behaviors in good or bad performance. 
They adopted machine learning techniques to automatically infer failures by analyzing the correlations between performance and system components. 
Jiang et al.~\cite{jiang2017causes} proposed CAM to failure cause analysis for test alarm in system and integration testing. 
For a given test alarm, they searched the test logs of historical test alarms that may have the same failure cause with the new test log. 
In particular, the similar matching is conducted by using K nearest neighbors (KNN) algorithm between log vectors, where log vectors are built on test log terms extracted by term frequency-inverse document frequency (TF-IDF). 
Following that, Amar et al.~\cite{amar2019mining} extended CAM by removing log lines that passed the test while keeping log lines only occurred in failed test logs. 
Then the historical logs were vectorized by a modified Line-IDF metrics. 
The vectors were utilized to utilized to train an EKNN model to identify most probable log lines that led to the test failures. 
Even though the retrieval-based methods are proposed to predict the test failure causes, they performed relatively poor with respect to the minority test failure causes, because they heavily rely on the sizes of the similar test logs, our approach based on heuristic rules can effectively handle the minority failure causes.

\subsection{Log-based Failure Prediction.}
Different from the root cause analysis which diagnoses the causes after the test fails, failure prediction aims to proactively predict the failure before it happens. 
Failure prediction is essential for predictive maintenance due to its ability to prevent failure occurrences and maintenance costs~\cite{abu2015failure}. 

A common practice of failure prediction is analyzing the system logs, which record the system status, changes in configuration, operational maintenance, etc. The source of failure can be divided into two categories, homogeneous systems, and heterogeneous systems, and the mainstreaming failure prediction approaches of different categories are different~\cite{he2021survey}.

In homogeneous systems (e.g., large-scale supercomputers), failure prediction approaches mainly focus on modeling sequential information. Sahoo et al.~\cite{sahoo2003critical} collected the system log of components' health status and leveraged several time series models to predict the health of each node in the system through indication metrics, such as the percentage of system utilization, usage of network IO, and system idle time. 
%Russo et al.~\cite{} converted the log sequence into the vector and employed support vector machines to predict failure. 
Klinkenberg et al.~\cite{klinkenberg2017data} trained a binary classification model from the system log and detected the potential node failure given a time sequence of monitoring data collected from each node.
Das et al.~\cite{das2018desh} adopted deep learning technology and proposed Desh to predict the failure of each node. It firstly recognized the log events chain leading to node failure, then it trained the log events chain recognition with expected lead times to node failure. Finally, Desh can predict the lead time of specific node failure.

In heterogeneous systems (e.g., cloud systems), failure prediction approaches mainly focus on modeling relationships among multiple components.
Chen et al.~\cite{chen2019outage} proposed AirAlert to find the dependence between the alerting signal extracted from system logs by the Bayesian network. Then AirAlert predicted failure based on a gradient boosting tree.
Lin et al.~\cite{lin2018predicting} designed MING which combined the LSTM and Random Forest model to o find the relationship between logs and the failure from temporal and spatial features.

\section{Threats to Validity}
\label{sec:threats}
Several threats to validity are related to our research: 
\noindent\textbf{Threats to internal validity} are related to potential errors in the code implementation and experimental settings. 
To reduce the errors in automatic evaluation, we have double checked the code of our approach and baselines.
Regarding the experiment results, we have carefully tuned the parameters of baseline approaches and used them in their highest performing settings for comparison.

\noindent\textbf{Threats to external validity} are related to the generalizability of the our experimental results. 
The generalizability of the root cause prediction algorithm in {\sc NCChecker} should be further explored, since our algorithm may be sensitive to datasets. 
To alleviate this threat, we evaluate our approach over industry datasets with more than 10K test logs. 
The ground truth of our dataset are of high quality because they are manually labelled by software developers/testers.

\noindent\textbf{Threats to construct validity} relate to suitability of our evaluation metric selection. 
We use the widely-accepted evaluation metrics (i.e., Precision/Recall/F1-score) to evaluate the effectiveness of our approach and baselines in our experiments. 
Since our dataset is highly imbalanced with respect to different types of causes, we use macro Precision/Recall/F1 metrics to estimate the overall performance. 
In addition, there are other software artifacts such as source/test code we did not use in this study, the performance of our approach may be further improved by leveraging more software artifacts.

\section{Conclusion}
\label{sec:con}
This research aims to automatically predict test failure causes for failed test logs. 
To address this task, we first collected more than 10K test logs from our industry partner and manually labeled the failed reason for each failed test log. 
We propose an approach named {\sc NCChecker} (Naive Failure Cause Checker) by leveraging log parsing and heuristic rules. 
Extensive experiments have demonstrated its effectiveness and promising performance for test failure causes prediction. 
% Our model is quite efficient and memory saving. 
Considering the effectiveness and simplicity of our approach, we recommend relevant practitioners to adopt our approach as a baseline for the failure causes prediction task.

\section*{ACKNOWLEDGMENT}
\label{sec:ack}
This research is supported by the Starry Night Science Fund of Zhejiang University Shanghai Institute for Advanced Study, Grant No. SN-ZJU-SIAS-001. 
This research is partially supported by the Shanghai Sailing Program (23YF1446900) and the National Science Foundation of China (No. 62202341). 
This research is partially supported by the Ningbo Natural Science Foundation (No. 2023J292). 
This research was also supported by the advanced computing resources provided by the Supercomputing Center of Hangzhou City University. 
The authors would like to thank the reviewers for their insightful and constructive feedback.

% \balance
% \bibliographystyle{ACM-Reference-Format}
% \bibliography{samples}

\balance
\bibliographystyle{ACM-Reference-Format}
\bibliography{samples}

\end{document}